# To the low-temperature technologies methodology: the clean superconductor free energy fluctuations calculation in the micro- and macrostructures descriptions of superconductor

Iogann Tolbatov

*Physics and Engineering Department, Kuban State University, Krasnodar, Russia*

(talbot1038@mail.ru)

The Ginzburg - Landau theory is used for the superconducting structures free energy fluctuations study. On its basis, we have defined the value of the heat capacity jump in the macroscopic zero-dimensional sample and in the zero-dimensional microstructures ensemble of the total volume equal to the macroscopic sample volume. The inference is made that in the Ginzburg - Landau methodology frameworks, it is essential to take into account the superconducting clean sample effective dimensionality only on the last stage of its thermodynamical characteristics calculation.

The properties of the being studied nowadays high-temperature superconductors (organic, low dimensional, amorphous) differ strongly from those of the conventional materials. The appearance of superconducting fluctuations in these superconductors leads to the crucial change of their conductivity, heat capacity, diamagnetic susceptibility etc. The appropriate tool for the fluctuation phenomena investigation is the Ginzburg – Landau (GL) theory [1]. The free energy functional:

$$\Phi[\psi(r)] = F_N + \int dV \{a|\psi(r)|^2 + \frac{b}{2}|\psi(r)|^4 + \frac{1}{4m}|\nabla \psi(r)|^2\} \tag{1}$$

introducted in this theory became the superconductivity phenomenological theory fundament.

In the finite volume system, the fluctuations smear the jump in heat capacity. In the case of the small superconducting sample of the effective dimensionality $D=0$ with the characteristic size $d \ll \xi(T)$, the main contribution to the free energy fluctuation part is defined by the space independent mode $\Psi_0 = \Psi\sqrt{V}$ [2]:

$$Z_{(0)} = \int d^2\Psi_0 \exp\left(-\frac{\Phi[\psi_0]}{T}\right) = \sqrt{\frac{\pi^3 VT}{2b}} \exp(x^2)(1 - erf(x))\bigg|_{x=a\sqrt{\frac{V}{2bT}}}. \tag{2}$$

Beyond the critical region, where $Gi_{(0)} \ll \varepsilon \ll 1$, one can find the expression for the free energy:

$$F_{(0)} = -T \ln Z_{(0)} = -T \ln \frac{\pi}{\alpha\varepsilon} \tag{3}$$

by means of the asymptotic expression for the $erf(x)$ function.

The total free energy of a clean superconducting sample can be estimated as the free energy of one such specimen (3) multiplied by their whole number $N_{(D)} = V\xi^{-D}(T)$:

$$F_{(D)} = -TV\xi^{-D}(T) \ln \frac{\pi}{\alpha\varepsilon}. \tag{4}$$

The jump in heat capacity in the second type phase transition point is described by expression:

$$\Delta C = C_S - C_N = -\frac{1}{VT_C}\left(\frac{\partial^2 F}{\partial \varepsilon^2}\right). \tag{5}$$

Considering that $\varepsilon = \ln\frac{T}{T_C}$, $\xi(T) = \frac{\xi_c}{\sqrt{\varepsilon}}$, where $\xi_c$ is the fluctuation Cooper pair coherence length in a clean superconductor, and calculating the second derivative of (4), we get:

$$\delta C = \frac{1}{V} e^{\varepsilon} \xi_c^{-D} \varepsilon^{\frac{D}{2}-2}\left(\ln\frac{\pi}{\alpha\varepsilon}\left(\varepsilon^2 + D\varepsilon + \frac{D}{2}\left(\frac{D}{2}-1\right)\right) - 2\varepsilon - (D-1)\right). \tag{6}$$

In the temperature region considered in the GL theory, id est in the region, where $\varepsilon = \ln\frac{T}{T_C} \ll 1$ (where $\varepsilon$ is the reduced temperature), the fluctuation order parameter value is small enough for to neglect the fourth order parameter term over $\psi(r)$ in comparison to the quadratic contribution corresponding to it. After the order parameter expanding in the Fourier series, the GL functional becomes the independent fluctuation modes sum [2]:

$$\Phi[\psi_k] = F_N + \sum_k \{a + \frac{k^2}{4m}\}|\psi_k|^2 = F_N \alpha T_C \sum_k (\varepsilon + \xi^2 k^2)|\psi_k|^2, \qquad (7)$$

where $\psi_k = \frac{1}{\sqrt{V}}\int \psi(r)e^{-ikr}dV$, the summation is carried out over the reciprocal space vectors $k$ (fluctuation modes). In the sample with the sizes $L_x$, $L_y$, $L_z$ $k_i L_i = 2\pi n_i$, the functional integral in the partition function is separated to a product of Gaussian type integrals over the modes mentioned above:

$$Z = \prod_k \int d^2\psi_k \exp\left(-\alpha\left(\varepsilon + \frac{k^2}{4m\alpha T_C}\right)|\psi_k|^2\right). \qquad (8)$$

Defining modes, we calculate the fluctuation contribution to free energy:

$$F(\varepsilon > 0) = -T\ln Z = -T\sum_k \ln\frac{\pi}{\alpha\left(\varepsilon + \frac{k^2}{4m\alpha T_c}\right)}. \qquad (9)$$

Hence, we find the fluctuation correction to heat capacity in the arbitrary dimension case:

$$\delta C_+ = \vartheta_D \frac{V_D}{V}\frac{(4m\alpha T_C)^{\frac{D}{2}}}{\varepsilon^{2-\frac{D}{2}}}, \qquad (10)$$

considering that $V_D = V, S, L, 1$ for $D = 3, 2, 1, 0$; where $\vartheta_D = \dfrac{\Gamma\left(2-\dfrac{D}{2}\right)}{2^D \pi^{\frac{D}{2}}}$ ($\Gamma(x)$ is the Euler gamma-function); $4m\alpha T_C = \xi_c^{-2}$.

While considering that the effective dimensionality equals zero, we assume that the two expressions (6) and (10) are not equivalent because the logarithm

$\ln\dfrac{\pi}{\alpha\varepsilon}$ in (6) tends to infinity as $\alpha_D = \dfrac{2\pi^2 DT_C}{7\zeta(3)E_F}$, where the effective dimensionality equals zero.

The carried out analysis shows that the macroscopic clean superconductor of the effective dimensionality $D = 0$ is not equivalent for the multitude of the same volume of clean microscopic structures of the zero effective dimensionality. That is why, in the GL formalism frameworks, the effective dimensionality is to be taken into account only at the last stage of its thermodynamical characteristics calculation.